\documentclass[journal,comsoc]{IEEEtran}
\usepackage[T1]{fontenc}
\ifCLASSINFOpdf
   \usepackage[pdftex]{graphicx}
   \DeclareGraphicsExtensions{.pdf,.jpeg,.png}
\else
\fi
\usepackage{amsmath}
\usepackage[cmintegrals]{newtxmath}
\begin{document}
%
\title{Key Enabling Technologies for Secure and Scalable Future Fog-IoT Architecture: A Survey}

%
%
%

\author{Jianli~Pan,~\IEEEmembership{Member,~IEEE}
		Yuanni~Liu,~\IEEEmembership{Member,~IEEE}
		Jianyu~Wang,~\IEEEmembership{Member,~IEEE}
		Austin~Hester,~\IEEEmembership{Member,~IEEE}
\thanks{First manuscript: March 1st, 2018.}
\thanks{J.~Pan, J.~Wang, and A. Hester are with the Department of Mathematics and Computer Science in University of Missouri, St. Louis, MO 63121, USA. (Email: pan@umsl.edu).}
\thanks{Y. Liu is with the Institute of Future Network Technologies, Chong Qing University of Posts and Telecommunications, China. (Email: liuyn@cqupt.edu.cn).}
}
\maketitle

\begin{abstract}
Fog or Edge computing has recently attracted broad attention from both industry and academia. It is deemed as a paradigm shift from the current centralized cloud computing model and could potentially bring a ``Fog-IoT'' architecture that would significantly benefit the future ubiquitous Internet of Things (IoT) systems and applications. However, it takes a series of key enabling technologies including emerging technologies to realize such a vision. In this article, we will survey these key enabling technologies with specific focuses on security and scalability, which are two very important and much-needed characteristics for future large-scale deployment. We aim to draw an overall big picture of the future for the research and development in these areas.  
 
\end{abstract}


%
\IEEEpeerreviewmaketitle

\section{Introduction and Overview of Future Fog-IoT Architecture} \label{sec:intro}

It is widely believed that a new paradigm of Internet of Things (IoT) is coming in which billions of IoT devices would connect to the Internet and bring many benefits to every aspect of human society. It could potentially enable smart home, smart transportation, smart health, smart energy and smart city applications. The wide deployment of these IoT devices and applications potentially would also bring significant challenges to the current centralized cloud computing model because of the massive amount of data generated in high speeds and the short latency requirement from some of the applications~\cite{panedge18}. Fog (or Edge) computing is deemed as a novel decentralized model to tackle the challenges. By providing local data storage, computation and networking in Fog or Edge nodes, the backbone Internet traffic burden could be alleviated, and the high-bandwidth and low-latency user experience of some IoT applications could be significantly improved. 

\begin{figure}[!t]
\centering
\includegraphics[width=\linewidth,keepaspectratio=true]{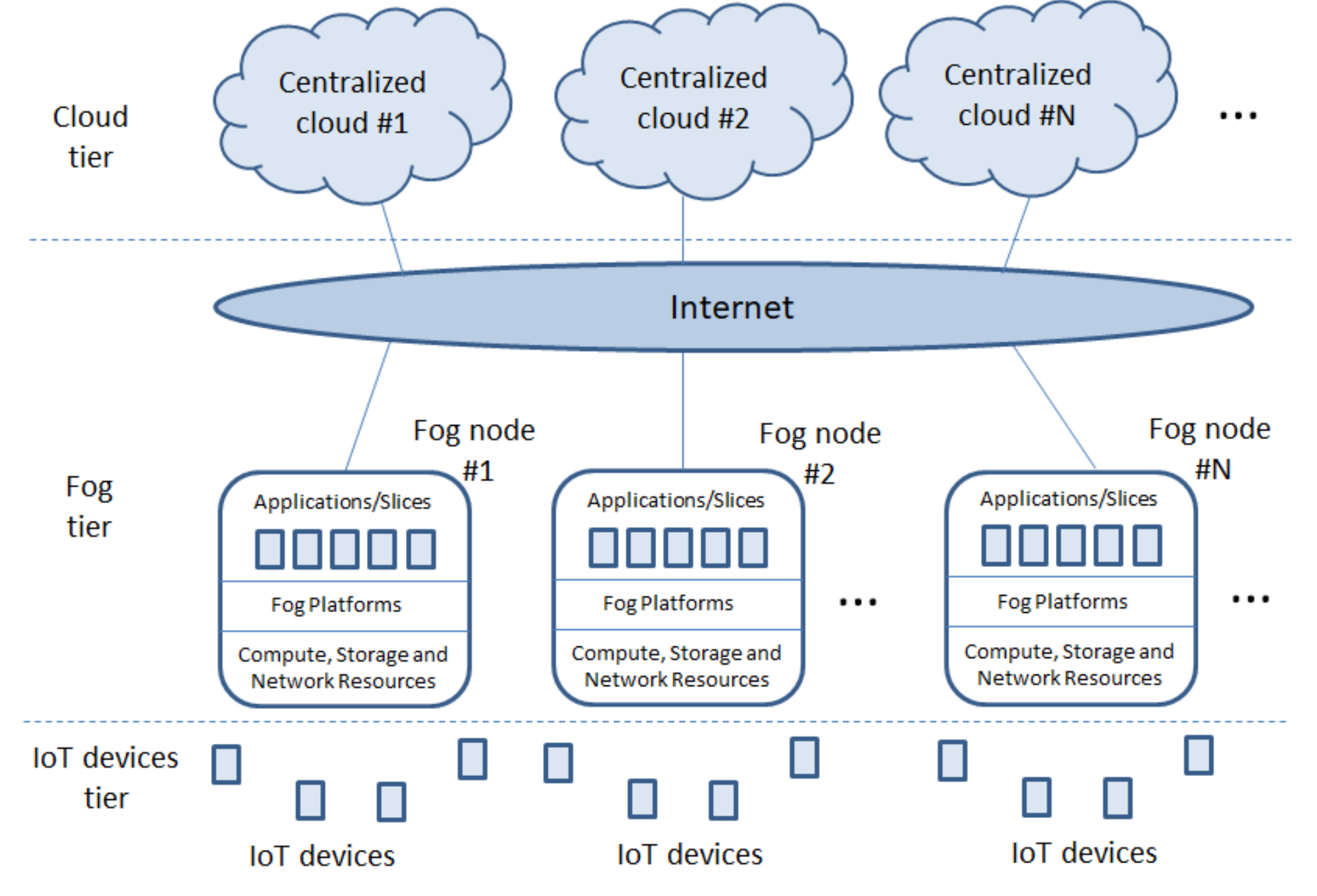}
\caption{Multi-tier Cloud-Fog-IoT framework.}
\label{fig:framework}
\end{figure}

A typical multi-tier framework involving centralized clouds, Fog nodes and IoT devices is shown in Fig.~\ref{fig:framework}. The cloud tier consists of traditionally centralized datacenters such as Amazon EC2 and Google Cloud. The IoT devices work with the Fog nodes locally first before sending only a limited size of data (if necessary at all) to the centralized clouds. The Fog tier includes various types of Fog platforms (CORD, Cloudlet, OpenStack, etc.) that virtualize compute, storage and networking resources, and create ``slices'' for various IoT applications that run on the shared Fog platforms. The Fog nodes are essentially small-sized shared datacenters in wired Internet or wireless telecommunications networks such as 5G. The Fog nodes can work in a standalone style for smart home applications, or in a decentralized way for smart community or smart city applications. 

However, to turn such a vision into reality, a secure and scalable ``Fog-IoT'' architecture is indispensible to potentially nurture a large variety of smart IoT applications for wide deployment in the future. A series of key enabling technologies including some emerging ones are needed to turn this vision into reality. These technologies are supposed to offer important functions from different angles for the ``Fog-IoT'' architecture. In this article, instead of surveying various ongoing Fog or Edge computing projects from industry or academia~\cite{panedge18}, we take a different approach, i.e., we focus on the key enabling technologies and their inner relationship, and discuss in detail how a future ``Fog-IoT'' architecture could be built to enable secure and scalable future IoT systems and applications. For simplification, we consider two major function groups: security and scalability, and the corresponding enabling technologies. 

\textbf{(1) Security group}. This group consists of functions such as privacy, confidentiality, transaction validity, traceability, tampering resistance, accountability, network isolation, and active defense. In this article, we consider four key technologies that could contribute to these new functions for the Fog-IoT architecture: (a) Blockchain and smart contracts; (b) multi-layer identities and naming other than IP; (c) Artificial Intelligence (AI) and machine learning (ML) technologies; (d) lightweight IoT security, and deception based active cyber defense technology.




\textbf{(2) Scalability group}. This group aims at enabling Fog-IoT systems and services more efficiently in a large scale. It includes functions such as shared Fog (Edge) infrastructure through application ``slices'', virtualized resources at Fog (Edge), deep programmability, lower costs, more dynamic, flexibility in scaling up or down, easier configuration and management, and more optimal and efficient coordination between Fog nodes and IoT devices. In this paper, we consider five types of key technologies contributing to these new functions: (a) Blockchain and smart contracts; (b) integration of Network Function Virtualization (NFV)~\cite{nfv14} and Software Defined Networking (SDN)~\cite{sdn14}; (c) orchestration, resource allocation, and onloading/offloading technologies; (d) global infrastructures such as GENI~\cite{geni14} and Planetlab for large-scale Fog-IoT experimentation; (e) AI and ML technologies.


\begin{figure}[!t]
\centering
\includegraphics[width=\linewidth,keepaspectratio=true]{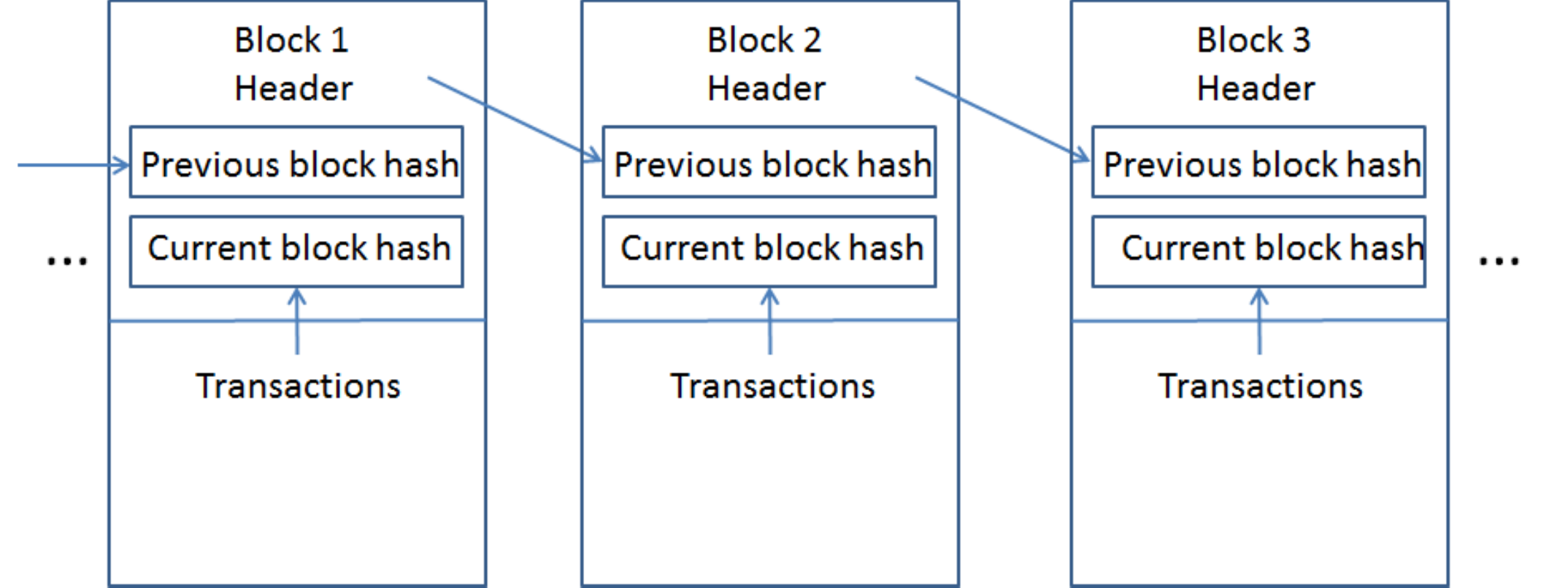}
\caption{A simple Blockchain structure.}
\label{fig:blockchain}
\end{figure}

Table~\ref{tab:functions} at the end of this article provides a complete view. The rest of this paper is organized as follows. From Section \ref{sec:blockchain} to Section~\ref{sec:other}, we discuss the key enabling technologies and their potential impacts to the future Fog-IoT architecture. Conclusions follow in Section~\ref{sec:conclusion}.



\section{Blockchain and Smart Contracts} \label{sec:blockchain}

In this section, we focus on Blockchain and smart contracts technologies that have both security and scalability group functions.

\subsection{Concepts of Blockchain and Smart Contracts}
Blockchain~\cite{bitcoin08} is an emerging technology that uses~\textit{decentralized and open ledgers} to record and maintain transactions in a verifiable and permanent way. It runs in a peer-to-peer overlay network in which every user can record transactions and validate other users' transactions. A very simple example of Blockchain structure is illustrated in Fig.~\ref{fig:blockchain}. The chain structure of the ledgers and the usage of hash function make it almost impossible to tamper the recorded data in the chain. It provides benefits in data/transactions persistence, validity, distributed storage, anonymity, privacy, traceability, tampering resistance and immediacy. Blockchain essentially eliminates the central authority and the users can transact with each other directly and securely without trusting each other. 


Originally introduced in 1994~\cite{smartcontract96}, a smart contract is defined as a computer program executed in a secure environment that directly controls digital assets. The emerging Blockchain technology offers a decentralized and secure environment for smart contracts to run. It enables automated processing, trust reduction (Blockchain is a trustless framework), and unambiguous enforcement (terms clearly expressed in code). Ethereum~\cite{ethereum14} is the first Blockchain-based smart contract platform. It specifies expressive programming language to enable much more applications than only transferring coins like in Bitcoin network. Transactions in Ethereum can be either normal transactions like Bitcoin transactions or transactions about smart contracts. 


\subsection{Blockchain and Smart Contracts for Fog-IoT Architecture}


\textbf{(1) For distributed Fog nodes}. The decentralized Blockchain matches well with the transiting trend from centralized cloud computing to decentralized Fog computing. It is promising to run Blockchain among distributed Fog nodes to facilitate \textit{secure data sharing, tracking and validation} for large-scale IoT applications. Fog nodes are usually resource-rich and they can run ``mining'' software to create new blocks and validate existing blocks to maintain the integrity and growth of the chain. With Blockchain, it is also possible to enable distributed data storage among Fog nodes for \textit{redundancy or fault tolerance} and it will be more difficult for the hackers to cause any significant disruption. Dynamic coordination among Fog nodes may also involve resource borrowing and lending with automatic payment as incentives for efficient resource usage. Such scenarios can be implemented with smart contracts among Fog nodes very conveniently, securely, and automatically. In addition, with Blockchain and smart contracts, the operational costs could be significantly reduced comparing with the centralized commercial clouds. 

\textbf{(2) For IoT devices and Fog node interaction}. Blockchain and smart contracts are very suitable to solve many problems with the IoT systems. For example, IoT applications usually rely on centralized servers for data storage, authentication, authorization, etc. The servers become single points of failure and hacking targets. With Blockchain, devices can directly transact with each other and significantly reduce the server loads and delay. It can accommodate even larger number of IoT devices and \textit{scalability will also be improved}. Transactions among IoT devices, and between IoT devices and Fog nodes can be facilitated by smart contracts with reliable, automatic and efficient executions. For example, smart contracts could enable escrow service for exchange, ``multisig'' service, digital wills, betting, prediction, insurance and micro-payment for computational services. Some of them can be very useful for the Fog-IoT environment. Furthermore, with Blockchain and smart contracts, IoT devices do not have to trust each other and significant cost can be saved that was otherwise spent in establishing additional trust relationship with each other. It is especially meaningful for resource-poor IoT devices.


\textbf{(3) For IoT cyberattacks resistence}. Blockchain and smart contracts could also help Fog nodes and IoT devices to detect and resist attacks. IoT devices usually have limited resources or capabilities to run full-scale security mechanisms to cope with attacks. This had been proved by the large-scale DDoS attack~\cite{ddos16}. Lightweight security methods could help but are not mature yet. With Blockchain and the decentralized ledgers, it is much more difficult for the hackers to break the Blockchain without enough CPU power to outpace the combined CPU power in the whole network and without being detected. Smart contracts also allow the IoT devices to define agreements on specific actions, behaviors and corresponding outcomes so that abnormal behaviors by the hackers can be automatically identified, detected and reported. Moreover, with the trustless Blockchain, it is even possible to implement a \textit{``zero-trust''} policy~\cite{zerotrust16} in the Fog network which could log all the transactions in the networks and help detect suspicious behaviors, potential misuses or attacks. Lateral movements from the hackers to the attacking targets will be much more difficult without being detected and blocked. 

\subsection{Blockchain and Smart Contracts Costs and Challenges}

The Blockchain benefits are not for free. The first problem is storage. The Blockchain ledgers need to be stored in the IoT devices and the size can increase as time passes. Computationally intensive ``mining'' should be done by Fog nodes instead of the IoT devices. Blockchain operation could create additional overhead traffic which may be undesirable for bandwidth-constrained situations. Ethereum Blockchain has also privacy issues when more data become public.

\section{Integration of NFV and SDN for Fog-IoT} \label{sec:nfvsdn}

This section focuses on the integration of NFV~\cite{nfv14} and SDN~\cite{sdn14} for scalability group function.

\subsection{Virtualization and Sharing via NFV}

NFV~\cite{nfv14} uses virtualization technologies to create network node functions and building blocks that could connect and chain together for various networking services. NFV eliminates the necessity of using traditional solely-purposed and expensive hardware for each network function. Instead, various Virtualized Network Functions (VNFs) can be dynamically and flexibly launched on industry standard high-volume servers, switches and storage devices for different networking functions. NFV facilitates sharing of computation, storage and networking resources while the actual virtualization can be implemented in hardware, OS, desktop, application and network level. Various containers and Virtual Machines (VMs) based virtualization technologies are very commonly seen.  

With NFV in the Fog-IoT architecture, the Fog nodes essentially form an edge cloud computing platform with virtualization and sharing capability. Various containers and VMs can be launched and configured into VNFs and chained for specific functionalities or even for an individual Fog-IoT application. In such an architecture, each IoT application is delivered and deployed as an independent ``slice'' over the same physical Fog infrastructure. Multi-application coexistence and running without interference also becomes possible. At last, NFV also leads to great improvement of agility and reduce the capital expenditure (CAPEX) and operation expenditure (OPEX). 

\subsection{Deep Programmability via SDN}

SDN~\cite{sdn14} is defined as the separation of data plane from control plane and using a centralized network intelligence and control instead of the traditionally static and decentralized control for more efficient networking. Deep programmability in network configuration, control, management, monitoring and troubleshooting are also enabled by SDN. The upper-level applications and networks services were abstracted from the underlying infrastructure. Typical examples of SDN protocols include ONF's OpenFlow, Cisco's Open Network Environment, and Nicira's Network Virtualization Platform.  

The deep programmability enabled by SDN is very useful for the future Fog-IoT architecture. When deploying IoT applications, from the Fog side, the applications may involve a series of VNFs launched from one Fog node or multiple decentralized Fog nodes. They need to be programmed, configured and chained effectively by SDN to deliver a new IoT application over the shared Fog infrastructure. SDN essentially creates network abstractions to allow increased flexibility and application-aware behaviors for various Fog-IoT applications.

\subsection{NFV and SDN Integration}

NFV and SDN are not necessarily dependent on each other. Instead, they are quite complementary and suitable to work together for Fog-IoT environment~\cite{panedge18}. Using NFV at the Fog nodes, various VNFs using heterogeneous hypervisors, containers, or VMs can be launched flexibly and chained dynamically for various IoT applications. Via these VNFs, computation, networking and storage resources can be allocated and utilized much more efficiently. For IoT applications, these VNFs also need deep programmability, control agility, easy configuration and management that are provided by SDN. Because of the respective benefits of NFV and SDN, and their complementary essence, it is very promising to integrate these two enabling technologies for the future Fog-IoT architecture. An example of such an integration effort is in our previous work on ``HomeCloud''~\cite{ict16}.

\begin{figure}[!t]
\centering
\includegraphics[width=\linewidth,keepaspectratio=true]{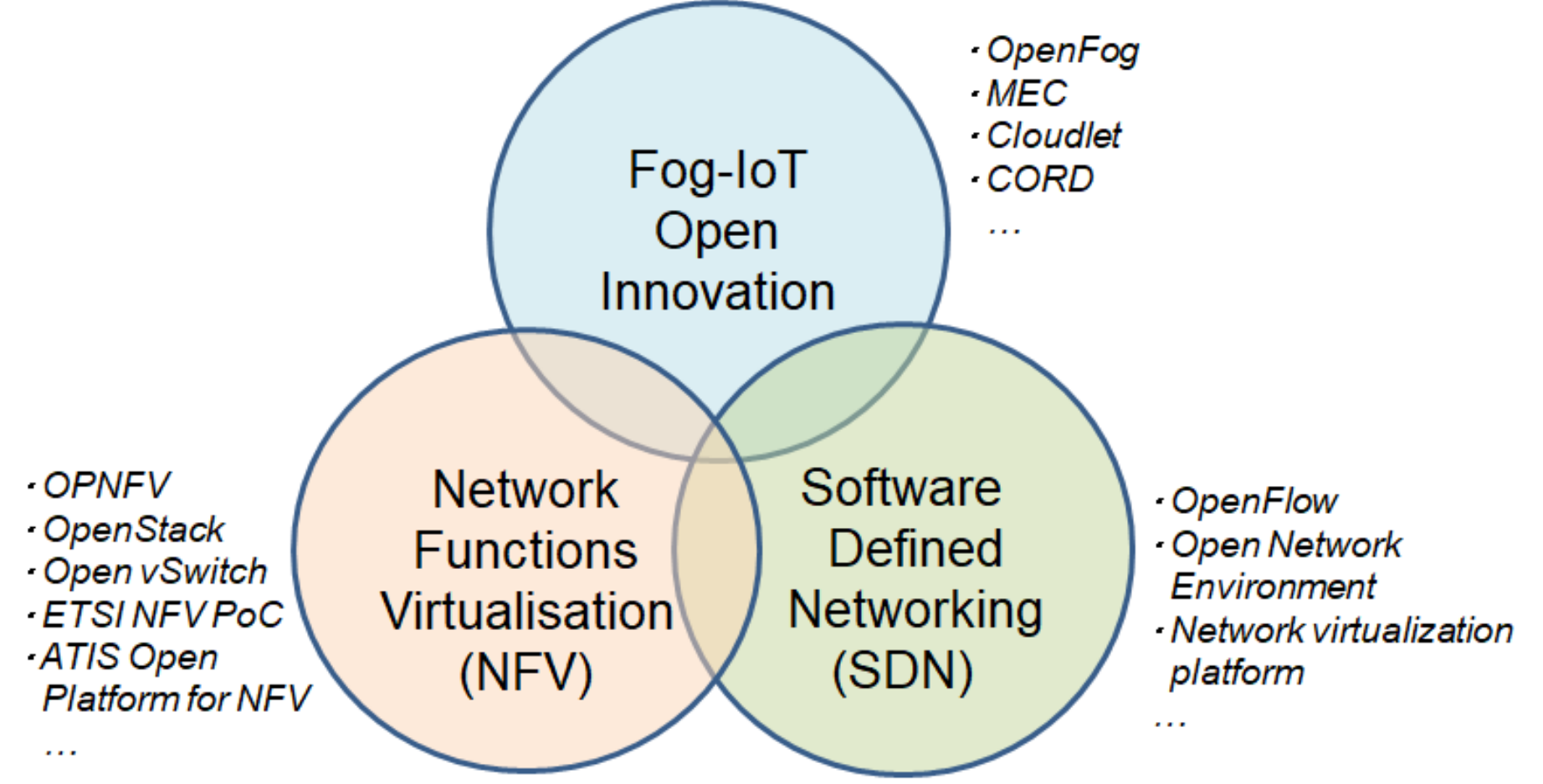}
\caption{Open Fog-IoT innovation based on NFV-SDN integration.}
\label{fig:nfvsdn}
\end{figure}

Although NFV and SDN are still evolving and developing respectively, both of them advocate open standards and open source innovations. We envision an open Fog-IoT innovation environment built on top of the coherent integration of NFV and SDN. Creating such an open eco-system is very important to break monopoly and nurture more innovations especially from small- and medium-sized innovators. We depict the relationship among different factors in such a new eco-system into the Fig.~\ref{fig:nfvsdn}. In the figure, we also list some related projects or efforts.
 
\section{Fog-IoT Orchestration, Resource Allocation, and Onloading/Offloading} \label{sec:orch}

In this section, we discuss a series of technologies for scalability group functions in Fog-IoT architecture.


\subsection{Automated Fog-IoT Orchestration}

\emph{Orchestration} is the set of operations or methods that the cloud providers use to manipulate and control the hardware and software resource usage for application delivery. It consists of all the tools for selection, deployment, monitoring and controlling resources during the whole life cycle of the applications. The state-of-the-art orchestration approaches are mostly based on the centralized cloud computing, and they are application-specific, highly customized, and often use manual methods or conditional checks (If-then-else) which increase complexity and are more error-prone.

For future Fog-IoT architecture, it is necessary to develop new automated tools with suitable abstractions to allow IoT applications to be deployed over decentralized Fog nodes, and to ensure that high-level demands are satisfied. As an example, an ``intent'' or ``goal'' oriented \textit{northbound approach} could be very useful for application-independent automated orchestration. This would solve the problems of the application-specific methods by the centralized clouds. Appropriate abstraction will eliminate the usage of manual and conditional checks, and to reduce errors. Also, automation would also be very useful to enable optimal resource allocation and provisioning in the Fog nodes. The Fog nodes can scale up or down the IoT applications and dynamically provision resources to achieve the optimal goals.  

\subsection{Optimal Resource Allocation}

Most of the future Fog-IoT applications would involve one or more decentralized Fog nodes. A large number of IoT devices may request computation, networking and storage assistance simultaneously. They be mobile and have significantly different demands in bandwidth, delay, or response time. To serve them well, Fog nodes have to be able to first monitor the applications and devices status, and then dynamically allocate resources among applications running on the shared Fog nodes via resource virtualization. In each of the application, resources also need to be allocated optimally to each user.

The Fog nodes may be subject to a series of constraints in computation power, memory, storage and energy. The IoT devices may have some dynamicity in sending bursty requests to Fog nodes, and also have constraints and requirements in bandwidth, delay and response time. When putting all these constraints together and considering decentralized Fog nodes coordination, optimal methods and algorithms are much needed to achieve high performance while making best use of the resources. Some typical decentralized optimization methods such as consensus algorithms, ADMM algorithm, Markov chain, game theory, and machine learning base approaches could be adopted to achieve the goals. 

\subsection{Fog-IoT Onloading or Offloading}

Onloading or offloading~\cite{offloading15} are typical coordination between the resource-rich Fog nodes (or edge cloud) and relatively resource-poor IoT devices when the IoT devices need assistance in computation, storage or other resources. Onloading usually means the Fog nodes initiate and coordinate the assistance while in offloading it is often initiated by the IoT devices. In some cases, it is also possible to allow the IoT devices to offload to each other (such as the ``FemtoCloud'' project). 

\begin{figure}[!t]
\centering
\includegraphics[width=\linewidth,keepaspectratio=true]{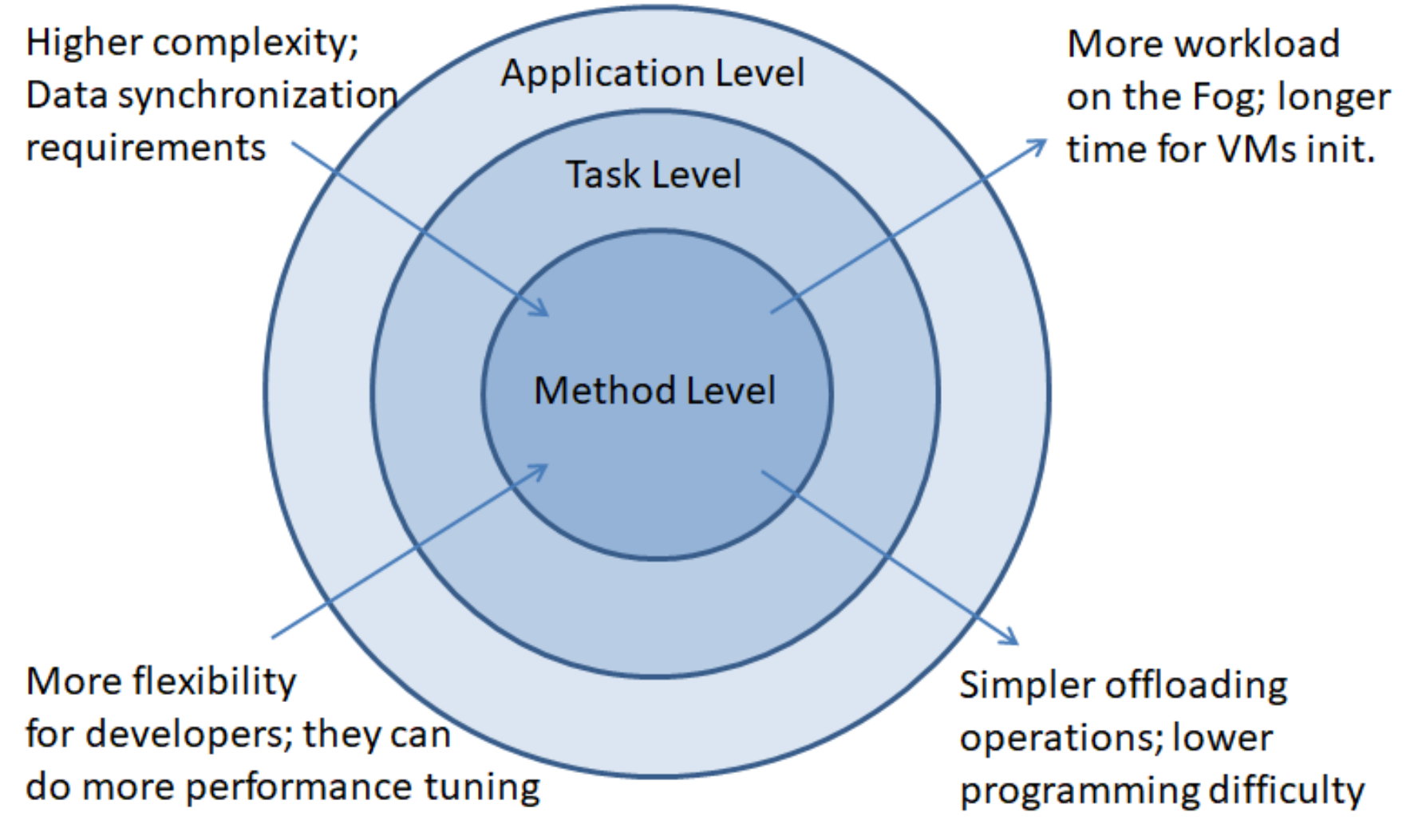}
\caption{Onloading/offloading granularities, and their advantages and disadvantages.}
\label{fig:offloading}
\end{figure}

Onloading or offloading can actually happen in different granularities: method level, task level, or application level depending on how to partition the programs into chunks to let them run either on Fog or on IoT devices. Methods are the smallest and they can be code fragments or functions. Tasks are larger than methods but smaller than applications. They are elements of an application that could be executed sequentially or in parallel. Applications level means that the whole software functions are executed in the Fog side (by VMs for example) and it is very thin in the mobile IoT devices side. These three methods have their advantages and disadvantages and we show them in Fig~\ref{fig:offloading}. 

Generally speaking, method-level onloading/offloading gives developers more flexibility to decide where the functions will be executed, but it suffers from high complexity and hard to do data synchronization between Fog nodes and IoT devices. Application-level onloading/offloading enjoys simpler operation and lower programming difficulty but it means more workload on Fog side including the time consumed for VMs initialization. For the future Fog-IoT architecture, we need to find a balance among these different methods to achieve the expected goals in performance and flexibility.

\section{Isolated Naming Space and Multi-layer Identity Bindings} \label{sec:identity}

In this section, we cover two technologies that are critical for both security and scalability group functions.


\subsection{Isolated Identity and Naming Space Other than IP}

Modern Internet uses IP as the universal addressing scheme. It enables rich connectivity and direct access between any heterogeneous hosts. This also brings much convenience for coordinated operations, maintenance and administrations of the decentralized devices. However, such design also brings significant threats and challenges to the ``weak'' and vulnerable IoT devices, especially for the devices in those critical industrial control systems such as smart grids or military defense systems. Traditional solutions usually deploy firewalls between the internal network and external Internet. But such firewalls enforce rules based on arbitrary IP addresses which can be spoofed or dynamically used by hackers to break through. Firewalls are also helpless inside the internal network when hackers move laterally toward the attacking targets.

To solve such problem, specifically for the future pervasive Fog-IoT environment, an isolated naming space other than IP is required to separate the identities of the devices from the corresponding addresses (like IP addresses). For example, instead of using the IP addresses as both identities and locators, we can design separate identifiers to be used for the internal IoT devices. External hosts cannot directly connect to an internal node except they use the identifiers to establish secure sessions (authenticating and exchanging security keys, etc.) and share a cryptographic binding before they can see each other. A typical example of such an additional identity and naming space is the ``host identifier'' that was used in the Host Identity Protocol (HIP)~\cite{hip08}. By doing this, it allows a certain degree of isolation and security among different portions of the networks that are not supposed to communicate directly with each other without appropriate security bindings being set up first. 

\subsection{Multi-layer Identities Dynamic Binding and Resolution}

\begin{figure}[!t]
\centering
\includegraphics[width=0.6\linewidth,keepaspectratio=true]{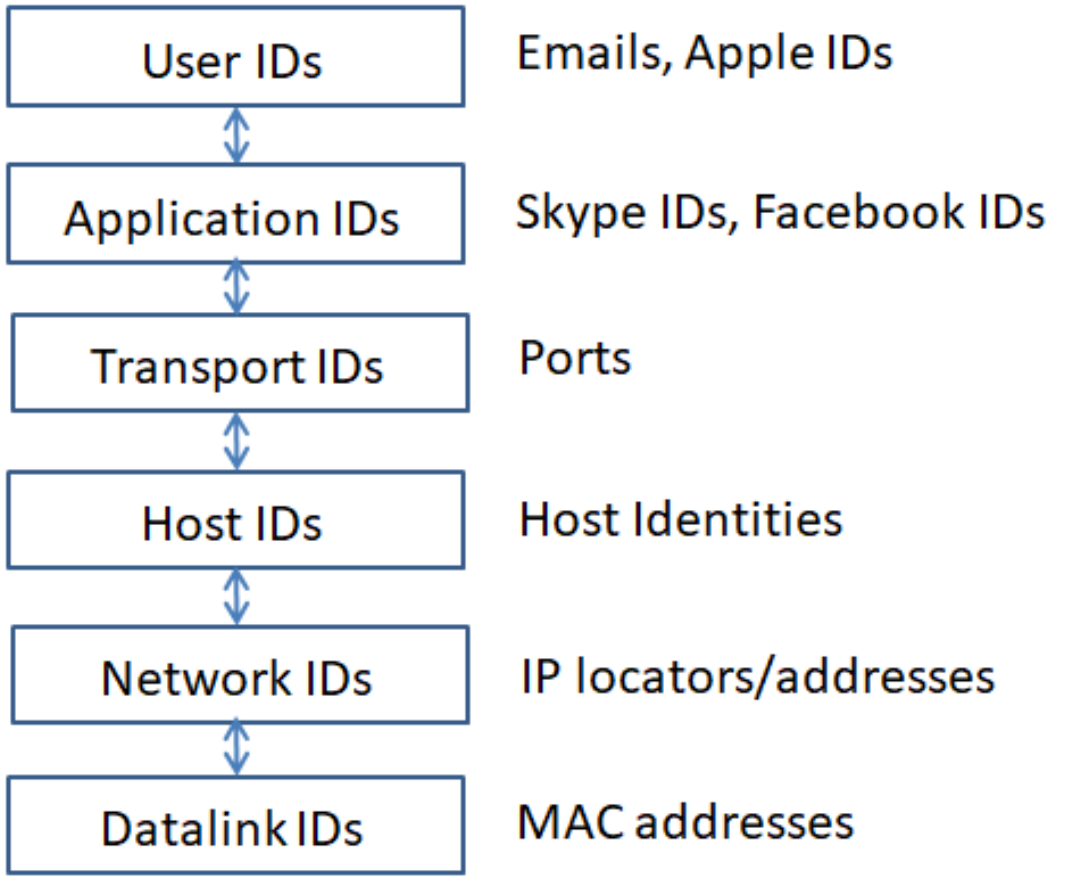}
\caption{Multi-layer identities and dynamic bindings.}
\label{fig:id}
\end{figure}

As shown in Fig.~\ref{fig:id}, the Internet protocol stack consists of multiple layers and each layer has its own identifiers (IDs). For example, the Apple IDs and skype IDs are the user-level identities associated to specific people and applications. Transport layer has port numbers associated to applications. Network layer has IP addresses as the identifiers as well as the locators. Data link layer has MAC addresses as the identifiers. However, these identifiers are used for different purposes and they have loose relationship except the double semantic underlying IP address which had been complained by many Internet researchers~\cite{jsac10}. Such a loose coupling among identifiers of different layers make it difficult to know the true identities of the actual users, especially when large-scale DDoS network attacks happened and many hosts were controlled by the hackers. There lack effective mechanisms to identify and control the users of many machines.

Decentralized systems for multi-layer identities' dynamic binding and resolution can be very useful for the security and scalability of future Fog-IoT architecture and applications. Such bindings can be used effectively in access control, which has been proven in the Google's tiered access control architecture~\cite{zerotrust16} adopting ``zero-trust'' security principles. They can also be useful to enable global scale high-level mobility (MobilityFirst project) or secure named data based networking (e.g., NDN project). For the Fog-IoT architecture specifically, the bindings among user IDs, application IDs, host IDs, and IP addresses can potentially be used instead of only IP addresses to uniquely identify a specific role in the Fog-IoT networks. For example, abnormal bindings can be identified when the hackers break in and try to pretend they are the legitimate users because they can hardly have the right multi-layer bindings to disguise themselves. It can be very useful to protect the users, applications, and infrastructure of the Fog-IoT environment.


\begin{table*}[!ht]  
\centering
\caption{Key enabling technologies and corresponding functions for Fog-IoT architecture.} 
\includegraphics[width=\linewidth,keepaspectratio=true]{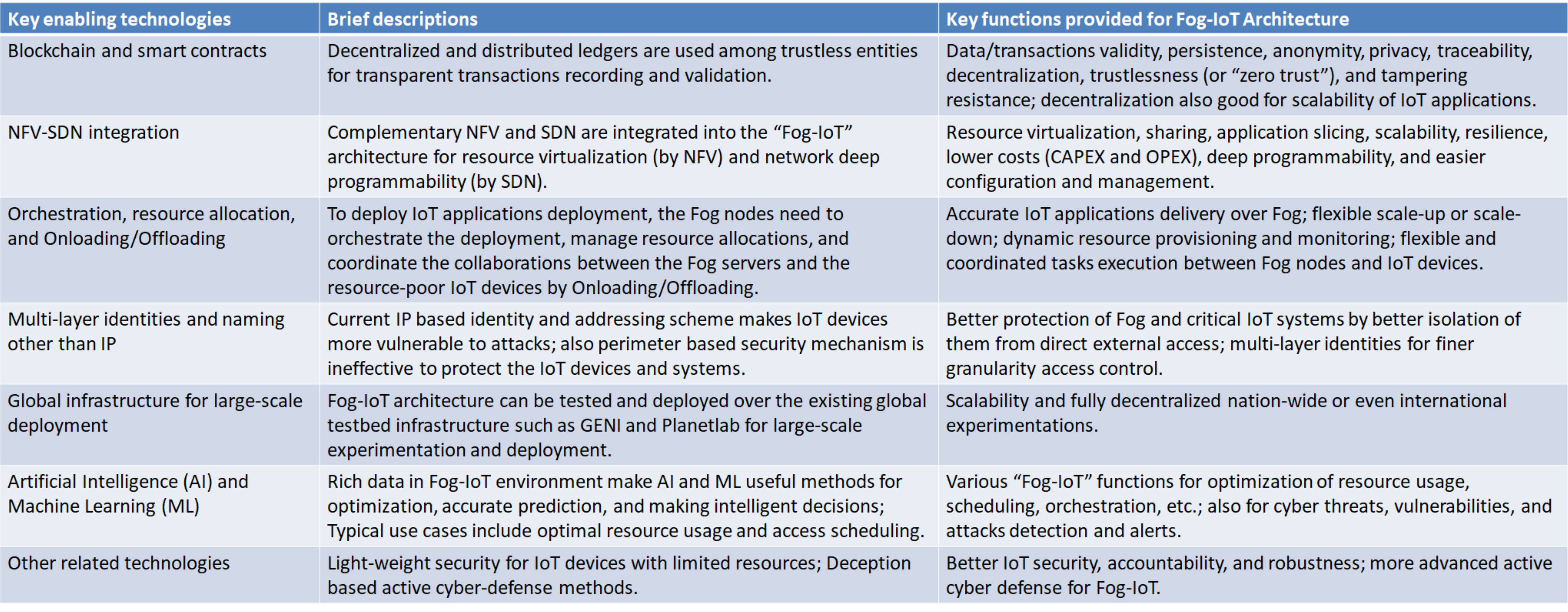}
\label{tab:functions}
\end{table*}

\section{AI and Machine Learning for Fog-IoT} \label{sec:aiml}

In recent years, AI and machine learning (ML) technologies (Deep Learning~\cite{deep15} in particular) have been widely used in various areas such as speech recognition, image processing and autonomous driving. In more and more occasions, applications powered by AI and ML can achieve better performance and results than human. When large datasets are available, AI and ML technologies powered by high-performance GPUs or other specialized chips can train the models, learn the patterns, make predictions and optimal decisions. In the future Fog-IoT architecture, there are three very useful places that AI and ML technologies can be useful:

\textbf{(1) Functional assistance in various Fog-IoT applications.} We know that Fog computing can be particularly beneficial to those IoT applications that need high bandwidth, low latency, and fast response. AI and ML technologies can be more effective for any ``human in the loop'' scenarios. Some typical such applications include Augmented Reality (AR)/Virtual Reality (VR), smart vehicles, face detection/recognition, autonomous driving, accessibility assistance, and smart health. AI and ML technologies can help make smart and optimal decisions based on the collected large datasets for specific IoT applications. 

\textbf{(2) Optimizing Fog and IoT interaction and coordination}. As we discussed in Section~\ref{sec:orch}, many interaction and coordination between Fog nodes, and between Fog and IoT devices need optimization methods and algorithms. Some typical operations include orchestration, resource allocation, and onloading/offloading. AI and ML technologies can be applied into most of these scenarios. For example, when many IoT devices using various applications request helps from a Fog node, AI and ML could potentially learn from the past patterns, predict the upcoming loads, and then reserve resources smartly in advance to accommodate bursty situations. 

\textbf{(3) Fog-IoT active security operations}. Most of the existing security mechanisms such as authentication, encryption and access control are passive solutions. With AI and ML technologies, active defense for Fog-IoT environment could be enabled for security related operations such as activities monitoring, misuses identification, threats and vulnerabilities detection, activity logging and automatic alarming or reacting. For example, AI and ML empowered Fog-IoT architecture could identify ongoing threats and attacks, raise alarms, and even take actions without human intervention. Some even believe it is possible to deploy ``robo-hunter'' empowered by AI to actively search in the network space for potential threats and breaches, and automatically resolve issues or raise alarms.

\section{Other Scalability and Security Related Enabling Technologies} \label{sec:other}

In this section, we discuss some other potential enabling technologies for both security and scalability. 

\subsection{Wide-area Deployment Using Global Infrastructures} \label{sec:wide}

Currently there are a series of nation-wide or even international networking and cloud computing infrastructures for research experimentation and education purposes. Typical examples include GENI~\cite{geni14} and PlanetLab from U.S., FIRE from Europe, VNode in Japan, and NICTA from Australia. Federation attempts were also made to interconnect these testbeds. The future Fog-IoT architecture can be deployed and tested on these testbeds for larger-scale experimentation. Security and scalability performance of the new architecture designs could have the opportunities to be validated and evaluated. Furthermore, it can also test the IoT applications running in large scales, and even see how the architecture and designs perform in situations involving multi-tenancy and multiple ``slices'' coexistence. 

\subsection{Lightweight Security for Fog-IoT}

Because of the limited resources and capabilities of many IoT devices, they cannot afford full-scale mechanisms in encryption, authentication and authorization. A tradeoff is that they can run lightweight instead of full-scale security algorithms and solutions. These lightweight methods are usually simplified versions while providing acceptable security that can satisfy the actual application requirements without significant sacrifice in security performance. With these methods, IoT devices can work longer with limited battery capacity while trying to work securely as well. However, more work still need to be done in this aspect, especially when there are still no clear standards to evaluate ``how secure is the lightweight solution to be considered secure enough''. Possibly a more thorough classifying framework is needed to define and categorize various IoT devices' capabilities in computation, networking and storage respectively, and to provide corresponding recommended lightweight security methods for different scenarios.

\subsection{Deception-based Fog-IoT Cyber Defense}

Deception-based defense is one of the active cyber defense methods that uses a series of ways to mislead, perplex, and capture malicious hackers. Typical methods includes address hopping, network telescopes and honeypots~\cite{iot16}. Deception-based defense can be very useful in the Fog-IoT environment. The Fog nodes are usually more powerful than the IoT devices and they can act as the active defenders by setting traps and honeypots to lure the attackers to use fake credentials. These attackers are trapped in the honeypots and are misled to take attacking actions in the defenders' favors or simply on the wrong targets. Since all the activities in the honeypot will be monitored and logged thoroughly, these data can be very useful for the attacking behaviors and pattern analysis. Generated insights can further help fortify the defending systems and mechanisms. However, the battle between the attackers and defenders is not easy to be ended, and we expect more advanced deception based cyber defense mechanisms to become available for the Fog-IoT environment in the future.

After we discussed the key technologies and their detailed functions for the Fog-IoT architecture, we here briefly summarize them in the Table~\ref{tab:functions}.

\section{Conclusions} \label{sec:conclusion}

Fog computing is a recent paradigm shift from the existing centralized cloud computing due to the trend of IoT. To enable Fog computing to work well with future ubiquitous IoT applications, a series of enabling technologies including some emerging technologies are needed. In this article, we envisioned a secure and scalable Fog-IoT architecture and surveyed the key enabling technologies. We aimed to draw big picture of the research and development in these areas.


%

%

\section*{Acknowledgment}

The work is supported in part by National Security Agency (NSA) under grants No.: H98230-17-1-0393 and H98230-17-1-0352, and by National Aeronautics and Space Administration (NASA) EPSCoR Missouri RID research grant under No.: NNX15AK38A.

\ifCLASSOPTIONcaptionsoff
  \newpage
\fi



%

%


\begin{IEEEbiography}[{\includegraphics[width=1in,height=1.25in,clip,keepaspectratio]{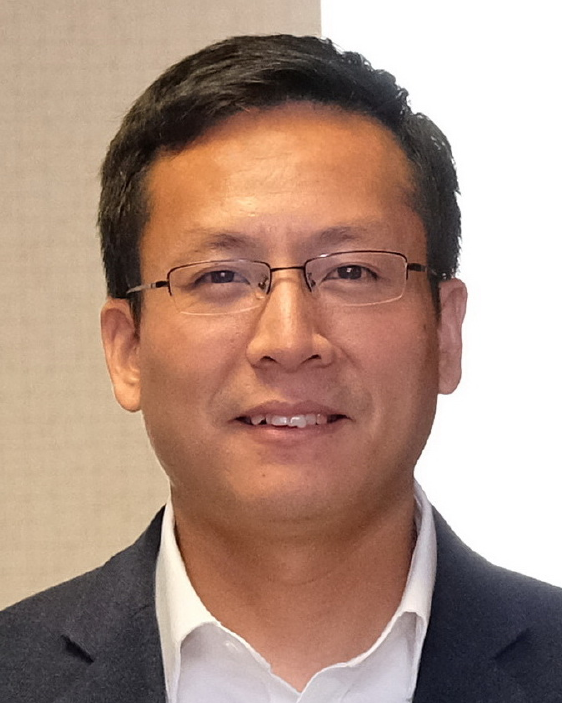}}]{Jianli Pan}

is currently an Assistant Professor in the Department of
Mathematics and Computer Science at the University of Missouri,
St. Louis. He obtained his Ph.D. degree from the Department of
Computer Science and Engineering of Washington University in
St. Louis. He also holds a M.S. degree in Computer Engineering
from Washington University in Saint Louis and a M.S. degree in
Information Engineering from Beijing University of Posts and
Telecommunications (BUPT), China. He is currently an associate
editor for both IEEE Communication Magazine and IEEE Access. His
current research interests include edge clouds, Internet of Things (IoT), Cybersecurity, Network Function Virtualization (NFV), and smart energy. 
\end{IEEEbiography}

\begin{IEEEbiography}[{\includegraphics[width=1in,height=1.25in,clip,keepaspectratio]{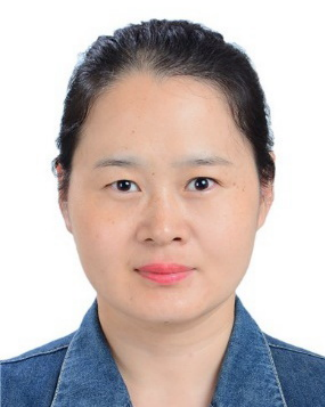}}]{Yuanni Liu}
is an associate professor at the Institute of Future Network Technologies, Chong Qing University of Posts and Telecommunications. She received her Ph.D. from the Department of network technology Institute, Beijing University of Posts and Telecommunications, China, in 2011. Her research interests include mobile crowd sensing，IoT security, and data virtualization.
\end{IEEEbiography}

\begin{IEEEbiography}[{\includegraphics[width=1in,height=1.25in,clip,keepaspectratio]{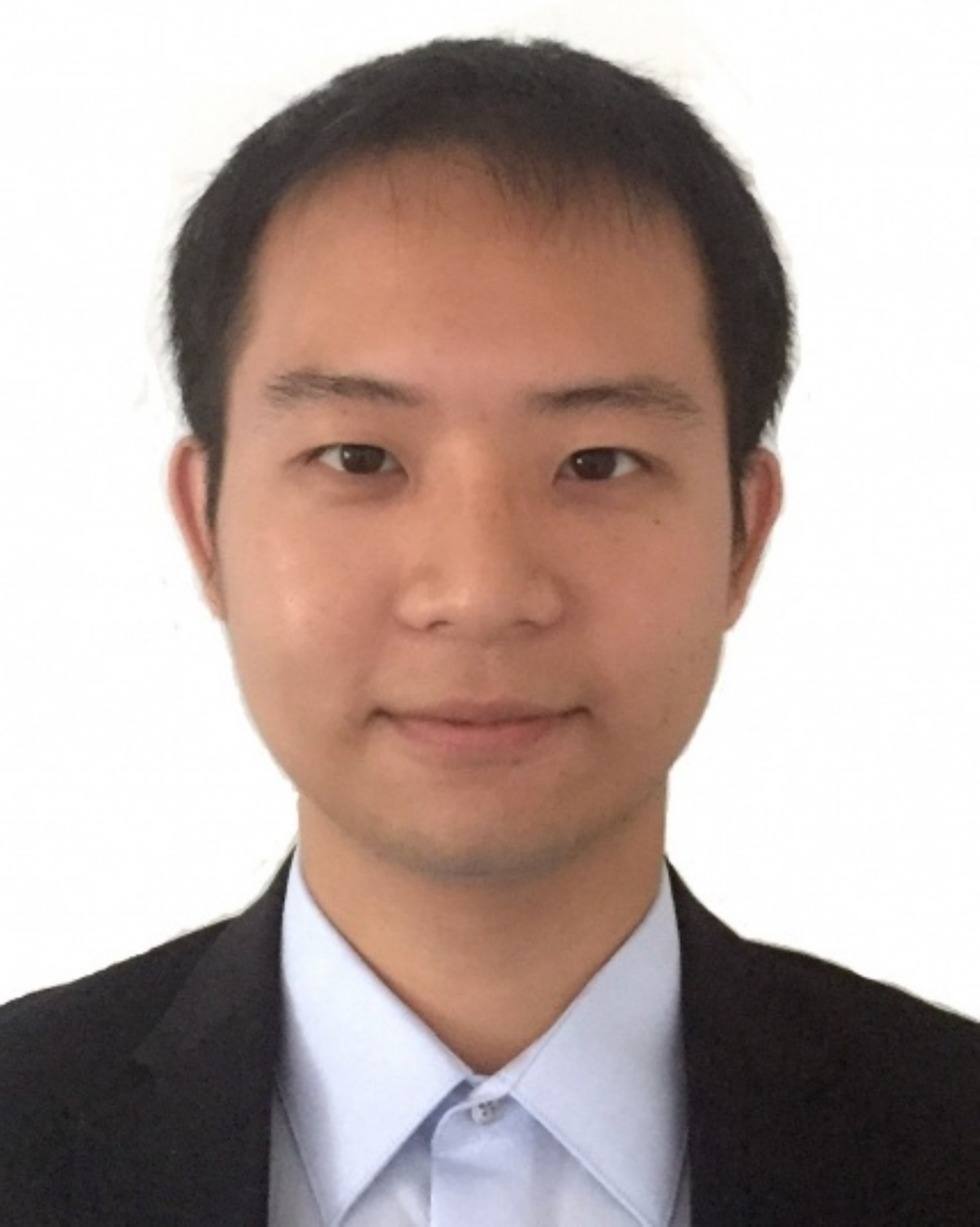}}]{Jianyu Wang}
is currently a Ph.D. student with the Department of Mathematics and Computer Science at the University of Missouri, St. Louis. He received an M.S. in Electrical and Computer Engineering from the Rutgers University, New Brunswick. His current research interests include edge cloud and mobile cloud computing. 
\end{IEEEbiography}

\begin{IEEEbiography}[{\includegraphics[width=1in,height=1.25in,clip,keepaspectratio]{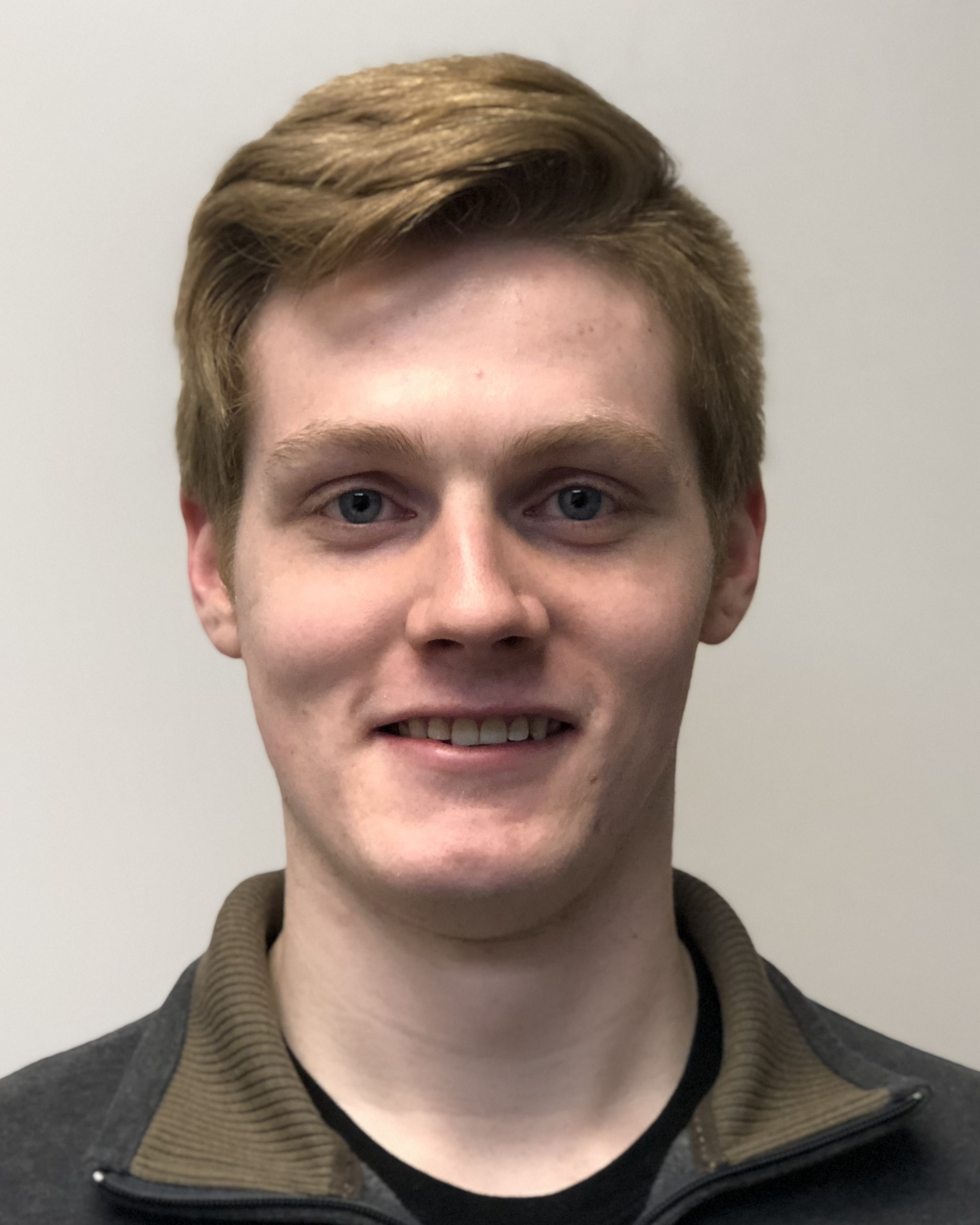}}]{Austin Hester}
is currently an undergraduate student with the Department of Mathematics and Computer Science at the University of Missouri, St. Louis. His current research interests include Internet of Things and Blockchain.
\end{IEEEbiography}




\vfill


\end{document}